\documentclass{article}
\usepackage{ijcai25}
\usepackage{amssymb} 
\usepackage{times}
\usepackage{soul}
\usepackage{url}
\usepackage[utf8]{inputenc}
\usepackage[small]{caption}
\usepackage{graphicx}
\usepackage{amsmath}
\usepackage{amsthm}
\usepackage{booktabs}
\usepackage[switch]{lineno}
\usepackage[hidelinks]{hyperref}
\usepackage{algorithm}
\usepackage[table]{xcolor}  
\usepackage{algpseudocode}%
\usepackage{float}
\usepackage{newfloat}
\usepackage{multirow}
\usepackage{makecell}
\usepackage{xcolor}         
\usepackage{colortbl}
\usepackage{dsfont}
\usepackage{amsmath}
\usepackage{bbding}
\usepackage{enumitem}
\usepackage{booktabs}

\usepackage{titlesec}

\newcommand*{\adaptor}{corrected adaptor}

\title{Device-Cloud Collaborative Correction for On-Device Recommendation}

\def\correspond{%
    \ifnum\value{eqfn}=0%
    \footnote{Corresponding authors.}%
    \setcounter{eqfn}{\value{footnote}}%
    \else%
    \footnotemark[\value{eqfn}]%
  \fi%
}%

\author{
    Tianyu Zhan$^1$\and
    Shengyu Zhang$^1$\footnotemark[1]\and
    Zheqi Lv$^1$\footnote{Corresponding authors.}\and
    Jieming Zhu$^2$\and
    Jiwei Li$^1$\and
    Fan Wu$^3$\And
    Fei Wu$^1$ \\
    \affiliations
    $^1$ Zhejiang University \\
    $^2$ Huawei Noah’s Ark Lab \\
    $^3$ Shanghai Jiao Tong University \\
    \emails
    \{yuzt,sy\_zhang,zheqilv,wufei\}@zju.edu.cn,
    jiemingzhu@ieee.org,
    bdlijiwei@gmail.com,
    fwu@cs.sjtu.edu.cn\\
}

\begin{document}

\maketitle
\begin{abstract}
With the rapid development of recommendation models and device computing power, device-based recommendation has become an important research area due to its better real-time performance and privacy protection. Previously, Transformer-based sequential recommendation models have been widely applied in this field because they outperform Recurrent Neural Network (RNN)-based recommendation models in terms of performance. However, as the length of interaction sequences increases, Transformer-based models introduce significantly more space and computational overhead compared to RNN-based models, posing challenges for device-based recommendation. To balance real-time performance and high performance on devices, we propose Device-Cloud \underline{Co}llaborative \underline{Corr}ection Framework for On-Device \underline{Rec}ommendation (CoCorrRec). CoCorrRec uses a self-correction network (SCN) to correct parameters with extremely low time cost. By updating model parameters during testing based on the input token, it achieves performance comparable to current optimal but more complex Transformer-based models. Furthermore, to prevent SCN from overfitting, we design a global correction network (GCN) that processes hidden states uploaded from devices and provides a global correction solution. Extensive experiments on multiple datasets show that CoCorrRec outperforms existing Transformer-based and RNN-based device recommendation models in terms of performance, with fewer parameters and lower FLOPs, thereby achieving a balance between real-time performance and high efficiency. Code is available at \url{https://github.com/Yuzt-zju/CoCorrRec}.
\end{abstract}
\section{Introduction}
With the rapid development of recommendation models and device computing power, deploying recommendation models on devices has become feasible. Recommendation on devices, being closer to the data source compared to cloud-based recommendation, demonstrates better real-time performance and privacy. Additionally, device-based recommendation helps alleviate the computational pressure on the cloud, making it an important area of research~\cite{gong2020edgerec,ref:intelligent}.
Device-based recommendation is generally responsible for reranking the candidate list provided by the cloud model based on the user's real-time behavior sequence~\cite{ref:dccl,lv2025collaboration}. Therefore, device-based recommendation models are mainly temporal recommendation models. The mainstream device-based recommendation models are divided into Transformer-based temporal recommendation models~\cite{ref:SAMRec,ref:ALBERT4Rec_DeBERTa4Rec,ref:sasrec} and Recurrent Neural Network (RNN)-based temporal recommendation models~\cite{ref:gru4rec,ref:KrNN-P,ref:MrRNN}.

\begin{figure}[t]
    \centering
    \includegraphics[width=0.93\linewidth]{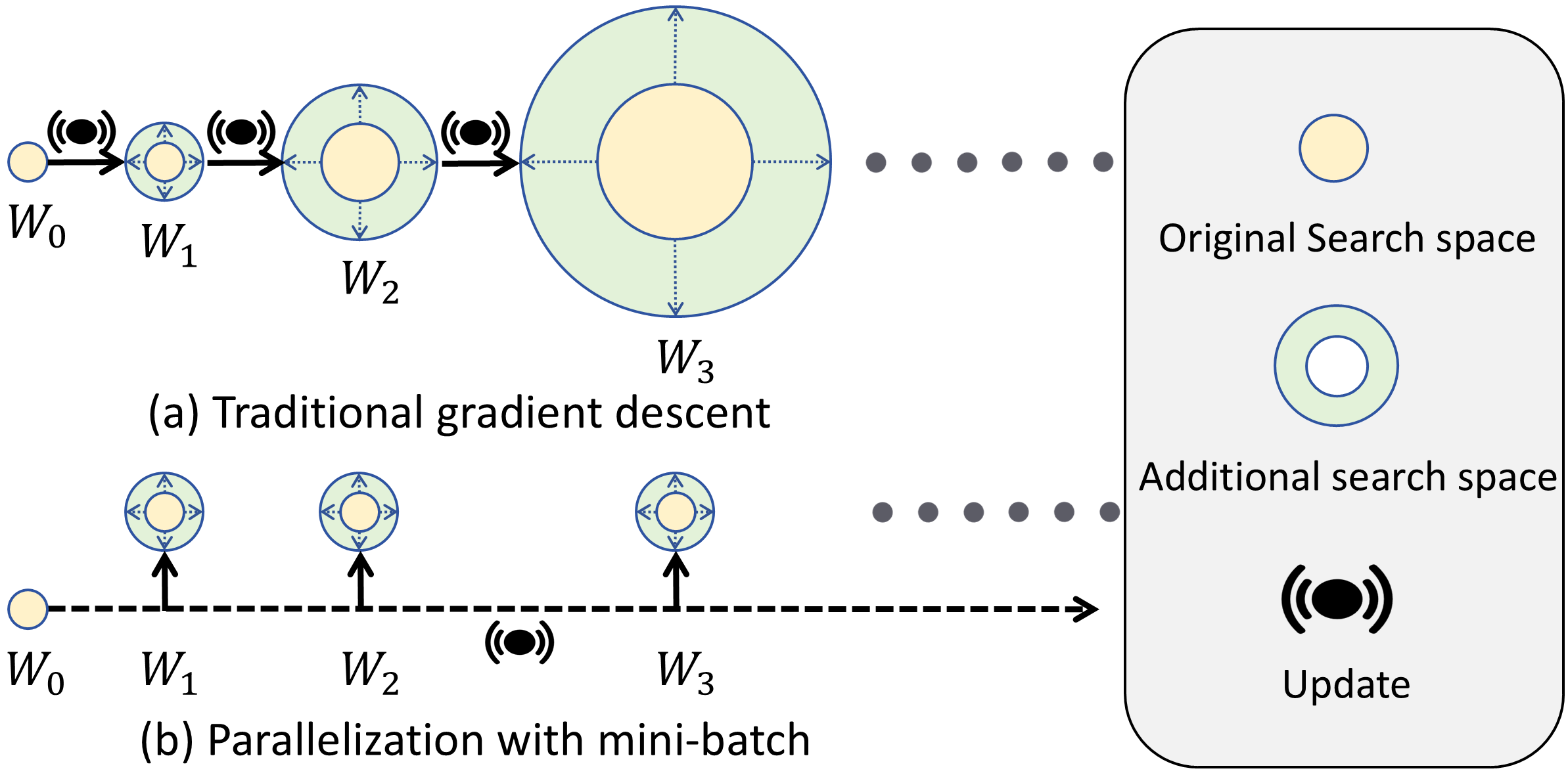}
    \caption{Parameter space variation in traditional gradient descent and mini-batch gradient descent. The size of the circle represents the size of the parameter search space.}
    \label{fig:search_space}
\end{figure}

Previously, Transformer-based models~\cite{ref:sasrec,ref:bert4rec,ref:mamba4rec} were widely applied in this field due to their superior performance compared to RNN-based recommendation models~\cite{ref:gru4rec}. The strong performance of Transformer-based models primarily stems from their ability to better retain early user behavior information when handling long sequences. In contrast, RNN-based models tend to gradually forget early user behavior information as the sequence length increases. However, the memory advantages of Transformer-based models come at a significant time and space cost. Specifically, assuming the length of the user behavior sequence is \(N\) and the tensor dimension is \(d\), ignoring the number of network layers, the time complexity and space complexity of Transformer-based models are \(O(N^2\cdot d)\) and \(O(N\cdot d)\), while for RNN-based models, they are \(O(N\cdot d)\) and \(O(d)\), respectively. Therefore, existing device recommendation models struggle to balance both real-time performance and high performance when handling long user behavior sequences.

To address the aforementioned challenges, we designed Device-Cloud \underline{Co}llaborative \underline{Corr}ection Framework for On-Device \underline{Rec}ommendation (CoCorrRec), which consists of two modules: the Self-Correction Network (SCN) and the Global Correction Network (GCN). After pretraining based on historical user behavior data, SCN and GCN are deployed on the device and the cloud, respectively. The goal of SCN is to retain the linear computational cost and fixed memory overhead of RNN-based models while improving their performance. To address the long-sequence forgetting issue in RNN-based models, we draw inspiration from test-time training and large language models, using self-supervised learning to dynamically update model parameters and create a hidden state model, enhancing the model's ability to remember long-sequence information~\cite{ref:TTT}. Considering real-time performance, SCN employs a mini-batch acceleration strategy to achieve parallel computation, enabling rapid and efficient updates to the device model.
Furthermore, we analyze the potential shortcomings of SCN: (1) SCN updates model parameters during the testing phase, which is key to reviving RNN-based models for device recommendations. However, the data distribution on the device is only a biased subset of the global distribution, which negatively impacts the model's generalization ability while enhancing its personalization. (2) The mini-batch acceleration strategy used by SCN processes multiple user behaviors simultaneously with shared hidden states. While this improves efficiency, it also reduces the effective search space (as shown in Figure \ref{fig:search_space}), limiting the model's ability to capture subtle patterns in user behavior. 
To address these limitations, we designed GCN, which receives the hidden state from the device's SCN, processed from the real-time user click sequences. GCN then performs device model parameter correction from a global perspective based on this hidden state. Since GCN uncovers similar user behaviors and trains using historical behavior data from numerous users, it exhibits significantly stronger generalization capabilities than the device model. As a result, CoCorrRec combines device-level personalization with cloud-level generalization, balancing real-time performance and high efficiency in device-based recommendations. Note that although GCN is designed to enhance SCN and can effectively improve performance, CoCorrRec does not rely on GCN. The device-based recommendation component of CoCorrRec, even when used on its own, can still achieve excellent performance. Extensive experiments on multiple datasets demonstrate that CoCorrRec outperforms existing device recommendation models, such as Transformer-based and RNN-based models, achieving performance improvements ranging from $0.06\%$ to $4.56\%$ across various metrics compared to the best-performing baseline models. Notably, CoCorrRec accomplishes these gains with fewer parameters and lower FLOPs, highlighting its ability to deliver superior recommendation accuracy while maintaining efficiency in terms of model size and computational cost.

The main contributions are summarized as follows:
\begin{itemize}
\item We introduce CoCorrRec, a novel recommendation system that integrates device-level personalization and cloud-level generalization to balance real-time performance and efficiency in device-based recommendations.
\item We design SCN, which utilizes self-supervised learning and mini-batch acceleration to improve the ability of RNN-based models to handle long user behavior sequences while keeping computational cost and memory overhead low.
\item We develop GCN, which corrects device model parameters using global user behavior patterns from the cloud, enhancing the model’s generalization ability and overall performance.
\item We conduct extensive experiments on multiple datasets, showing that CoCorrRec outperforms existing device recommendation models, such as Transformer-based and RNN-based models, with fewer parameters and lower computational overhead.
\end{itemize}

\section{Related Work}
\noindent\textbf{Sequential Recommendation.} Sequential recommendation models have traditionally relied on the Markov Chain assumption, where current interactions depend solely on the most recent interaction or a few preceding ones~\cite{ref:FPMC}. While these models are straightforward and effective, they fail to capture long-term sequential dependencies. To address this limitation, various neural networks, such as Recurrent Neural Networks (RNNs)~\cite{ref:gru4rec,ref:KrNN-P,ref:MrRNN}, Convolutional Neural Networks (CNNs)~\cite{ref:Caser}, Attention networks~\cite{ref:SAMRec,ref:ALBERT4Rec_DeBERTa4Rec,ref:STRec,su2023personalized,su2023enhancing} , Graph Neural Networks (GNNs)~\cite{ref:MSSG,ref:AutoSeqRec}, and Multi-Layer Perceptrons (MLPs)~\cite{ref:MLM4Rec,ref:FMLP-Rec}, have been employed to better utilize sequential information. More recently, advanced techniques such as contrastive learning and diffusion models~\cite{ref:diffusion4rec} have also been applied to sequential recommendation tasks.

\noindent\textbf{Device-Cloud Recommendation.} Device-cloud collaboration is an emerging approach in recommendation systems that combines the personalization capabilities of device-side models with the generalization power of cloud-based models~\cite{ref:device_cloud,ref:device_cloud_rec,ref:device_cloud_adarequest,fu2025forward,fu2024diet}. This approach addresses challenges inherent in traditional recommendation methods, such as user-oriented bias/fairness issues~\cite{ref:Fairness}, as well as privacy concerns~\cite{ref:fedfast}. Early efforts, including DCCL~\cite{ref:dccl} and MetaController~\cite{ref:MetaC}, developed collaborative frameworks where device-side models focused on personalization\cite{zhang2020devlbert} while cloud-side models provided global knowledge~\cite{ref:duet,ref:intelligent}. When the TTT model is deployed on a device, updating parameters during recommendations for a specific user resembles single-user fine-tuning, which may compromise the model's general knowledge. To mitigate this, the cloud-based network can provides similar user information to the specific model and perform parameter fusion\cite{Li2024MergeNetKM}, compensating for the loss of generalization.

\noindent\textbf{Test-Time Training.}
Test-Time Training (TTT) refers to a strategy where each test instance defines a unique learning problem, targeting its own generalization~\cite{ref:test}. As test instances lack labels, the learning problem must be framed using a self-supervised task. An early implementation of this concept, known as local learning, involves training on the neighbors of each test input before making predictions. This approach has been successfully applied across models~\cite{ref:test_LLM,zhan2025preliminary}. Previous research demonstrates that TTT with reconstruction notably enhances performance, particularly for outliers~\cite{ref:ttt-mae}. These improvements are even more pronounced when applied to video frames in a streaming context~\cite{ref:test_video}, where TTT operates autoregressively.


\section{Preliminary}
\subsection{Overview of Test-Time Training}
With the observation that self-supervised learning can compress a massive training set into the weights of a model such as LLMs, TTT layer represents a novel category of sequence modeling layers, where the hidden state is a model, and the update rule is defined as a step in self-supervised learning~\cite{ref:TTT}. Two primary instantiations of this class are TTT-Linear and TTT-MLP, in which the hidden state corresponds to a linear model and a two-layer MLP, respectively. In this work, we select TTT-Linear as the core component of the TTT Block.

For self-supervised learning in TTT layer, three low-rank projection learnable matrix $\theta_K$, $\theta_V$, $\theta_Q$ are introduced. Specifically, when processing a sequence $x_1$,...,$x_t$, $\theta_K$ applies corruption to $x_t$, $\theta_V$ extracts critical information for reconstruction and $\theta_Q$ ensures that $x_t$ remains aligned with the input dimensionality of $f$. The update rule for W is a gradient descent step: \begin{equation}
            W_t=W_{t-1}-\eta \nabla l(W_{t-1};x_t)
        \end{equation}
where the self-supervised loss function $l$ is:
        \begin{equation}
            l(W;x_t)=\|f(\theta_Kx_t;W)-\theta_Vx_t\|^2
        \end{equation}
            
The output token, $z_t$, is predicted for $x_t$ using $f$ with the updated weights $W_t$:
            \begin{equation}z_t=f(\theta_Qx_t;W_t)\end{equation}

\subsection{Mini-batch Strategy}
In naive TTT, each update of $W_t$ depends on the previous step $W_{t-1}$, which inhibits efficient parallel computation. The general gradient descent method can be expressed as:
\begin{equation}W_t=W_{t-1}-\eta \nabla l(W_{t-1};x_t)=W_0-\eta\sum_{s=1}^{t}{\nabla l(W_{s-1};x_s)}\end{equation}
To address this limitation, mini-batch gradient descent is employed. Let the TTT mini-batch size be $b$. $W$ can be computed in parallel as follows:
\begin{equation}W_t=W_{t'}-\eta\sum_{s=t'+1}^{t}{\nabla l(W_{t'};x_s)} \quad t'+1 \leq t \leq t'+b\end{equation}

where $ t' = t- \mod(t,b) $ is the last timestep of the previous mini-batch (or 0 for the first mini-batch). This approach enables parallel computation of $b$ gradient updates at a time.

\subsection{Dual Form}
\label{sec:dual_form}
To fully harness the matrix computing capabilities of the GPU, the calculation process for TTT has been optimized. Consider the simplest case of $l$, where $\theta_K = \theta_V = \theta_Q = I$, and focus on the first TTT mini-batch of size $b$. Additionally, assume that $f$ is a linear model. In this context, just compute the output $Z= [z_1,..., z_b]$ and initializing $W_b$ for the next mini-batch:
\begin{equation}
\left\{
\begin{aligned}
& W_b=W_0-2\eta{(W_0X-X)X^T} \\
& Z=W_0X-2\eta\Delta
\end{aligned}
\right.
\end{equation}
where 
\begin{equation}
    \Delta=mask(X^TX)(W_0X-X)
\end{equation}

\section{Methodology}
\begin{figure*}[!ht]
    \centering
    \includegraphics[width=\linewidth]{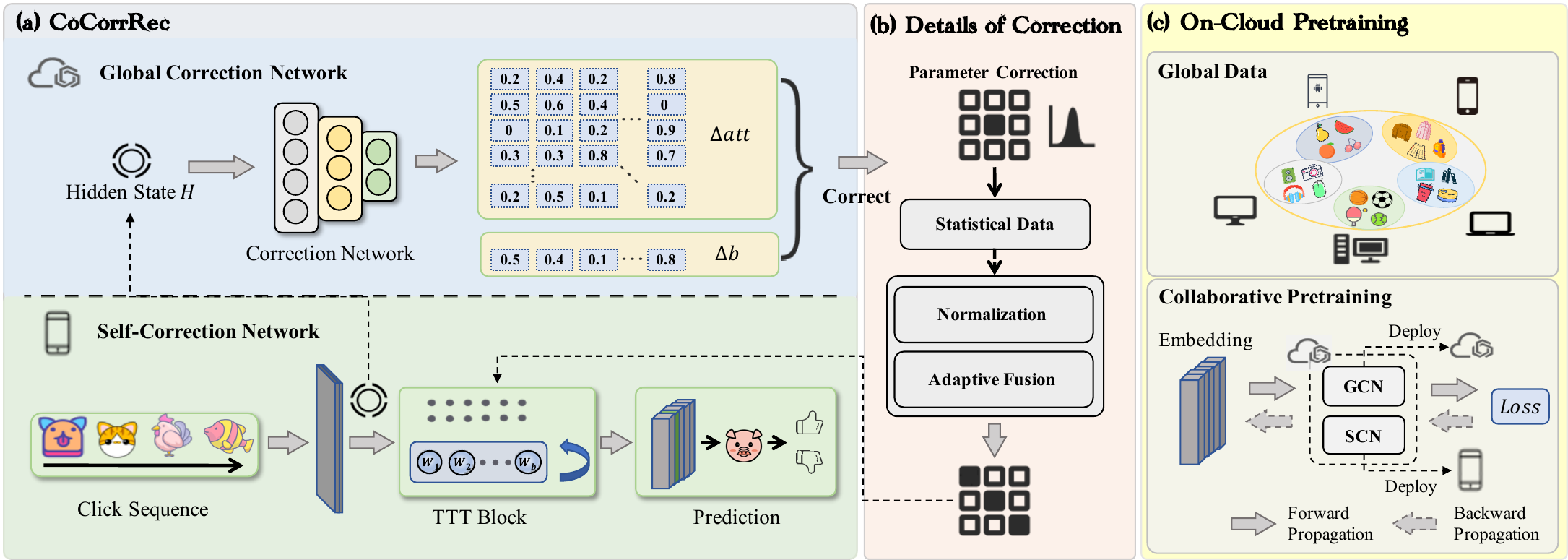}
    \caption{Overview of the CoCorrRec. (a) describes the pipeline of CoCorrRec inference, the on-device SCN uploads $H$ to the on-cloud GCN and subsequently downloads the correction generated by it. (b) describes the details of correction on the device. (c) describes the pretraining procedure on cloud via global data collected from massive devices.}
    \label{fig:main_pipline}
\end{figure*}
To enable recommendation models on devices to balance real-time performance and high performance, we designed a self-correction network based on TTT to improve RNN-based recommendation models.
It extract features from users' real-time click sequences, enabling high-quality recommendations with minimal resource usage. To further overcome two specific performance limitations of \adaptor{}, we developed a cloud-based computationally intensive  Correction Network (GCN). This network expands TTT's parameter search space and extracts knowledge from similar users, thereby integrating collaborative filtering information into the recommendation process.

\subsection{On-Device Self-Correction Network}\label{sec:TTT4Rec}
\begin{figure}[t]
    \centering
    \includegraphics[width=0.9\linewidth]{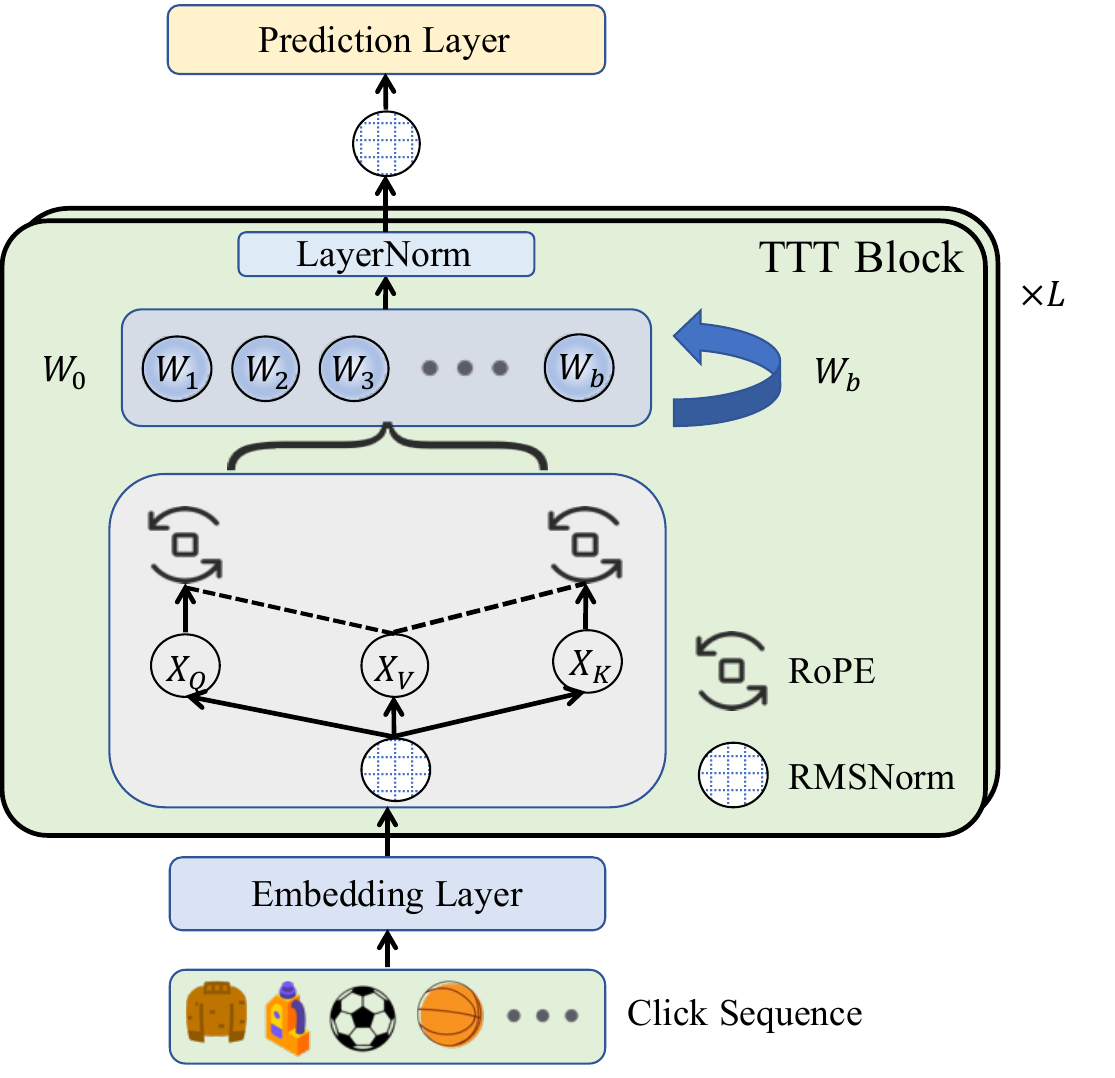}
    \caption{The overview of on-device SCN.}
    \label{fig:TTT(Base)}
\end{figure}

To ensure optimal efficiency and simplicity, the on-device SCN model $M(\cdot;\theta_M)$ is designed with a streamlined architecture, comprising only an embedding layer, TTT Block, and a prediction layer (as illustrated in a figure \ref{fig:TTT(Base)}).

The embedding layer maps item IDs to a high-dimensional space using a learnable embedding matrix, $E\in \mathbb{R}^{|V|\times D}$, where $D$ denotes the embedding dimension. When applied to the input item sequence $S_u$, we obtain the initial item embeddings $E_u$. To enhance robustness and mitigate overfitting, RMS normalization is applied after the embeddings are retrieved:
    \begin{equation}
        H=RMSNorm(E(S_u))
    \end{equation}
While the TTT block demonstrates potential as a sequence-to-sequence model, we focus on leveraging its strengths for sequential recommendation tasks. Based on practical application, the TTT block has been streamlined to further reduce computational cost and enhance the efficiency of the on-device model.
Starting with the hidden states $H$, a low-order projection is performed using $\theta_Q$, $\theta_K$, and $\theta_V$:
    \begin{equation}X_i=\theta_i H \quad i \in \{Q,K,V\}\end{equation}
Rotary positional encodings (RoPE) \cite{su2024roformer} are then applied to $X_Q$ and $X_K$ as follows, in which rotary embedding is calculated by $X_V$:
    \begin{equation}X_i=RoPE(X_i,X_V) \quad i \in \{Q,K\}\end{equation}
TTT Block employs self-supervised learning to process the input $X$. When using a mini-batch and dual form, the calculation of $W_b$ and the output $Z$ detailed in Algorithm \ref{alg:ttt} assuming that $\eta$ = 1 and the first mini-batch. Finally, it undergoes Layer Normalization (LN) and a final linear projection: 
\begin{equation}O=\theta_oLN(Z)\end{equation}
In the prediction layer, the final item embedding of O is used to compute the output prediction scores:
    \begin{equation}\hat{y}=Softmax({RMSNorm(O)}_{last}E^T) \in \mathbb{R}^{|V|}\end{equation}

\begin{algorithm}[!h]
\caption{TTT Block of SCN}
\label{alg:ttt}
\begin{algorithmic}
    \Statex \textbf{Input:} $H$
    \Statex \textbf{Output:} $W_b, Z$
    \Statex \textbf{Process:} $W_b, Z$
    \State \hspace{5mm} $X_Q,X_K,X_V \leftarrow \theta_Q H, \theta_K H, \theta_V H$
    \State \hspace{5mm} $X_Q,X_K \leftarrow \operatorname{RoPE}(X_Q,X_V),\operatorname{RoPE}(X_K,X_V)$
    \State \hspace{5mm} $\nabla_{X_K} l \leftarrow W_0 X_K - X_V$
    \State \hspace{5mm} $\nabla_{Z} l \leftarrow \sigma'(\nabla_{X_K} l)$ \\
    \hspace{5mm} \Comment{$\sigma'$ is the derivative of sigmod function}
    \State \hspace{5mm} $\nabla_{W_0} l \leftarrow \nabla_{Z} l (X_K)^T$
    \State \hspace{5mm} $W_b \leftarrow W_0 - \nabla_{W_0} l$
    \State \hspace{5mm} $b \leftarrow b_0-\text{mask}(\nabla_{Z} l)$
    \State \hspace{5mm} $attn \leftarrow \text{mask}(X_K^T X_Q)$
    \State \hspace{5mm} $Z \leftarrow W_0X_Q - \nabla_{Z} l \cdot attn + b$
    \State \Return $W_b, Z$
\end{algorithmic}
\end{algorithm}


\begin{table*}[!ht]

    \label{tab:main_table}
    \centering
    \renewcommand{\arraystretch}{1.05}
    \resizebox{0.92\textwidth}{!}{
    \begin{tabular}{c|c|c|c|c|c|c|c|c}
    \toprule[2pt]
    \multirow{2}{*}{\textbf{Dataset}} & \multirow{2}{*}{\textbf{Model}} & \multicolumn{7}{c}{\textbf{Metrics}} \\ \cline{3-9}
     &  & {UAUC} & {NDCG@5}  & {NDCG@10}  & {NDCG@20} & {HitRate@5}  & 
    {HitRate@10}  & {HitRate@20}  \\
    \midrule
    \midrule
    \multirow{7}{*}{\texttt{Beauty}} & {DIN} 
    & 0.7185 &	0.2525 & 0.2866 & 0.3102 & 0.3655 & 0.4703 & 0.5629 \\
    & {GRU4Rec} 
    & 0.7078  &	0.2577 & 0.2825 & 0.3103 & 0.3485 & 0.4254 & 0.5361  \\
    & {SASRec} 
    & 0.7314  &	0.2605 	&0.2920 & 0.3180 & 0.3929 & \underline{0.4852} & \underline{0.5876}  \\
    & {BERT4Rec}
    & 0.7178  &	0.2555 & 0.2829 & 0.3109 & 0.3445 & 0.4296 & 0.5403 \\
    & {Mamba4Rec}
    & 0.7220  &	\underline{0.3051} 	&\underline{0.3300} & \underline{0.3543} & 0.3871 & 0.4639 & 0.5603 \\
    \cline{2-9}
    & \cellcolor{blue!5}{CoCorrRec (w/o. GCN)} 
    & \cellcolor{blue!5}\underline{0.7331} & \cellcolor{blue!5}{0.2987} & \cellcolor{blue!5}{0.3256} & \cellcolor{blue!5}{0.3517} & \cellcolor{blue!5}\underline{0.4019} & \cellcolor{blue!5}{0.4858} & \cellcolor{blue!5}{0.5874} \\
    & \cellcolor{blue!5}CoCorrRec
    & \cellcolor{blue!5}\textbf{0.7342} 
    & \cellcolor{blue!5}\textbf{0.3126} 
    & \cellcolor{blue!5}\textbf{0.3371} 
    & \cellcolor{blue!5}\textbf{0.3626} 
    & \cellcolor{blue!5}\textbf{0.4108} 
    & \cellcolor{blue!5}\textbf{0.4867} 
    & \cellcolor{blue!5}\textbf{0.5888} \\
    
       \midrule
       \midrule
    \multirow{7}{*}{\texttt{Electronic}} & {DIN} 
    & 0.7991 & 0.3422 & 0.3799 & 0.4070 & 0.4596 & 0.5760 & 0.6829 \\
    & {GRU4Rec} 
    & 0.8257 &0.3528 & 0.3933 & 0.4237 & \underline{0.4762} & \underline{0.6016} & \underline{0.7218} \\
    & {SASRec} 
    & 0.8161 &0.3418  & 0.3820 & 0.4131 & 0.4639 & 0.5881 & 0.7107 \\
    & {BERT4Rec}
    & 0.8182 &0.3454 & 0.3851 & 0.4171 & 0.4665 & 0.5891 & 0.7156 \\
    & {Mamba4Rec}
    & 0.8198 & 0.3390 & 0.3787 & 0.4115 & 0.4632 & 0.5861 & 0.7157  \\
    \cline{2-9}
    & \cellcolor{blue!5}CoCorrRec (w/o. GCN)

    & \cellcolor{blue!5}\underline{0.8258}	
    & \cellcolor{blue!5}\underline{0.3550}	
    & \cellcolor{blue!5}\underline{0.3950}	
    & \cellcolor{blue!5}\underline{0.4258}	
    & \cellcolor{blue!5}{0.4754}	
    & \cellcolor{blue!5}{0.5989}	
    & \cellcolor{blue!5}{0.7210}	
    \\
    & \cellcolor{blue!5}CoCorrRec
    & \cellcolor{blue!5}\textbf{0.8315}	& \cellcolor{blue!5}\textbf{0.3629}	& \cellcolor{blue!5}\textbf{0.4021}	& \cellcolor{blue!5}\textbf{0.4327}	& \cellcolor{blue!5}\textbf{0.4862}	& \cellcolor{blue!5}\textbf{0.6074}	& \cellcolor{blue!5}\textbf{0.7282}	\\
    \hline
    \hline
    \multirow{7}{*}{\texttt{Yelp}} & {DIN} 
    & 
    0.9522 	& 0.5302 	& 0.5774 	& 0.5959 & 0.7333 & 0.8775 & 0.9497  \\
    & {GRU4Rec} 
    & 
    0.9522& 0.5247& 0.5724& 0.5916& 0.7318& 0.8778& \underline{0.9525}  \\
    & {SASRec} 
    & 
    0.9499& 0.5099& 0.5607& 0.5804& 0.7173& 0.8729& 0.9495 \\
    & {BERT4Rec}
    & \underline{0.9524} & \underline{0.5388} & 0.5830& 0.6000& \underline{0.7451} & \underline{0.8841} & 0.9514 \\
    & {Mamba4Rec}
    & 0.9514& 0.5301& 0.5776& 0.5954& 0.7387& 0.8832& 0.9528  \\

    \cline{2-9}
    & \cellcolor{blue!5}{CoCorrRec (w/o. GCN)} 
    & \cellcolor{blue!5}{0.9511}	
    & \cellcolor{blue!5}{0.5372}	
    & \cellcolor{blue!5}\underline{0.5835}	
    & \cellcolor{blue!5}\underline{0.6006}	
    & \cellcolor{blue!5}{0.7412}	
    & \cellcolor{blue!5}{0.8824}	
    & \cellcolor{blue!5}{0.9489}	\\
    
    & \cellcolor{blue!5}CoCorrRec
    & \cellcolor{blue!5}\textbf{0.9533}	
    & \cellcolor{blue!5}\textbf{0.5412}	
    & \cellcolor{blue!5}\textbf{0.5865}	
    & \cellcolor{blue!5}\textbf{0.6036}	
    & \cellcolor{blue!5}\textbf{0.7470}	
    & \cellcolor{blue!5}\textbf{0.8858}	
    & \cellcolor{blue!5}\textbf{0.9531}	\\
    \bottomrule[2pt]
    \end{tabular}
    }
\caption{Overall performance comparison between the baselines and CoCorrRec. The \textbf{best} performance is highlighted in bold and the second-best results are \underline{underlined}.}

\end{table*}
\subsection{On-Cloud Global Correction Network}
The On-Cloud Global Correction Network (GCN) generates dynamic parameter corrections for the \adaptor{} based on real-time samples from specific devices. Its purpose is to expand the search space and provide collaborative filtering information to enhance the recommendation process. Specifically, the Correction Network is implemented as a three-layer MLP, denoted as $C(\cdot;\theta_c)$ with parameters $\theta_c$. It takes $H$ from the device as input and predicts the correction values $K$. Due to its high computational cost (measured in FLOPs), the GCN is more suitable for cloud deployment rather than on-device implementation. Since it operates independently of the \adaptor{}, both can run in parallel.
Based on Algorithm 1, we instantiate GCN as $C_{attn}$ and $C_b$ to compute corrections for the key values $attn$ and $b$:
    \begin{equation}K_i = C_i(H;\theta_i) \quad i \in \{attn,b\}\end{equation}
After the cloud-generated results are produced, they are sent directly to the on-device model for correction. To ensure robust results,  $K_{attn}, K_b$ are then normalized to match the mean and variance of the corresponding values $attn$ and $b$:
    \begin{equation}K_i = \frac{K_i-mean(K_i)}{std(K_i)}*std(i)+mean(i) \quad i \in \{attn,b\}\end{equation}
To further refine the corrections, learnable fusion parameters $P$ and $Q$ $\in \mathbb{R}^{1\times 1}$ are introduced for adaptive fusion:

    \begin{equation}\left\{
\begin{aligned}
& attn = attn + P\cdot K_{attn} \\
& b = b + Q\cdot K_{b}
\end{aligned}
\right.\end{equation}

\subsection{Training and Inference}
The training of CoCorrRec is performed entirely in the cloud, utilizing abundant computational resources. All layers are optimized simultaneously using global historical data, $S_H=\{x_{H}^{(i)};y_{H}^{(i)}\}_{i=1}^{N_H}$, where ${N_H}$ represents the total number of training data samples. The loss function,  $\mathcal{L}_{total}$, is defined as follows:
    \begin{equation}\underset{{\theta_M,\theta_C}}{min\mathcal{L}_{total}}=\sum_{i=1}^{N_H}{D_{ce}(y_{H}^{(i)},M(x_{H}^{(i)};\theta_M,C(H_{x_{H}^{(i)}};\theta_C)))}\end{equation}
where $D_{ce}(;)$ denotes the cross-entropy between two probability distributions, and $H_x$ represents the hidden state of input $x$ as described in Section \ref{sec:TTT4Rec}.

Since the input $H_x$ to $C$ is computed before the TTT Block, the entire framework can correct the TTT Block during training and inference in parallel which is detailed in Section \ref{sec:time_delay}. This parallelization significantly improves the training and inference efficiency of CoCorrRec, making it more suitable for deployment in real-world environments.

After the CoCorrRec training is completed, the SCN component is deployed to specific devices for real-time re-ranking. Meanwhile, the computationally intensive GCN is retained in the cloud to enhance the recommendation performance of on-device SCN instance. For real-time samples, $S_R=\{x_R^{(i)}\}_{i=1}^{N_R}$, where $N_R$ represents the total number of real-time data samples, the recommendation process is formulated as:
    \begin{equation}\hat{y}_R^{(i)}=M(x_R^{(i)};\theta_M,C(H_{x_{R}^{(i)}};\theta_C))\end{equation}

\section{Experiments}
In this section, we conduct extensive experiments on three realworld datasets to answer the following research questions: 
\begin{itemize}
\item RQ1: What is the resource consumption required for various basic recommendation models (FLOPs and maximum memory usage)?
\item RQ2: What is the Time delay caused by device-cloud communication of the CoCorrRec?
\item RQ3: How does our CoCorrRec perform compared to the baselines under various experimental settings and how do the designs of it affect the performance?
\end{itemize}
\subsection{Experimental Settings}
\subsubsection{Datasets} We conduct experiments on three publicly available datasets in different scenarios. Amazon-beauty and Amazon-electronic are from an e-commerce platform Amazon\footnote{\url{https://jmcauley.ucsd.edu/data/amazon/}}\label{fn:amazon}, which covers various product categories such as books, electronics, home goods, and more. We also evaluated on Yelp\footnote{\url{https://www.yelp.com/dataset/}}\label{fn:yelp} which is a representative business dataset containing user reviews for different restaurants.
\begin{table}[ht]
\normalsize
\tabcolsep=5.5pt
  \centering
  \label{tab:dataset}
  \resizebox{0.46\textwidth}{!}{%
  \begin{tabular}{lcccc}
    \toprule
    \multicolumn{1}{l}{\textbf{Dataset}} & \multicolumn{1}{c}{\textbf{\# Users}} & \multicolumn{1}{c}{\textbf{\# Items}} & \multicolumn{1}{c}{\textbf{\# Interactions}} & \multicolumn{1}{c}{\textbf{Avg. Length}} \\ 
    \midrule
    Beauty & $4,255$ & $22,301$ & $96,053$  & $22.6$\\
    Electronic & $27,224$ & $75,749$ & $599,276$  & $22.0$\\
    Yelp & $18,046$ & $46,118$ & $1,358,413$ & $75.3$\\
  \bottomrule
\end{tabular}}
  \caption{Statistics of three datasets.}
\end{table}
For all datasets, we first sort all interactions chronologically according to the timestamps. And then discard users and items with interactions $\le$ 10. All user-item pairs in the dataset are treated as positive samples. In the training and test sets, the user-item pairs that do not exist in the dataset are sampled at 1:4 and 1:100, respectively, as negative samples.
The detailed statistics for each dataset after preprocessing are summarized in Table \ref{tab:dataset}.

\subsubsection{Baselines}
We compare CoCorrRec with several competitive baselines, including both RNN-based and Transformer-based methods. GRU4Rec \cite{ref:gru4rec} employs gated recurrent units (GRUs) to model behavior sequences, achieving excellent performance. DIN \cite{ref:din} introduces attention mechanisms to capture users’ interests related to the current product within their historical behavior. SASRec \cite{ref:sasrec} leverages self-attention to efficiently capture long-term dependencies in user behavior sequences. BERT4Rec \cite{ref:bert4rec} utilizes a bidirectional Transformer to extract contextual information from user behavior sequences, delivering more accurate recommendations. Lastly, Mamba4Rec \cite{ref:mamba4rec} highlights recent advances in state-space models (SSMs) for efficient sequential recommendation tasks.
\subsubsection{Evaluation} For performance comparison, we adopt three widely used metrics UAUC, HitRate@K and NDCG@K over the top-K items, where K is set as 5, 10 or 20. 
\begin{table}[!h]

    \label{tab:hyperparameters_training}
    \setlength{\arrayrulewidth}{0.5pt}
 \resizebox{0.48\textwidth}{!}{
    \begin{tabular}{c|c|c|c}
    \toprule[2pt]
    \textbf{Dataset} & \textbf{Model} & \textbf{Hyperparameter} & \textbf{Setting} \\ 
    \midrule
    \midrule
    \multirow{7}{*}{\makecell[c]{Beauty\\Electronic\\Yelp}} & \multirow{7}{*}{\makecell[c]{DIN\\GRU4Rec\\SASRec\\BERT4Rec\\Mamba4Rec\\CoCorrRec}} & GPU &  RTX-4090 GPU(24G) \\ 
    \multirow{7}{*}{} & & \cellcolor{gray!15}Optimizer & \cellcolor{gray!15}AdamW\\ 
    \multirow{7}{*}{} & & \makecell[c]{Learning rate} &  $\{1e^{-3}, 2e^{-3}, 5e^{-3}, 1e^{-2}\}$\\ 
    \multirow{7}{*}{} & & \cellcolor{gray!15}{Batch size} & \cellcolor{gray!15}1024 \\ 
    \multirow{7}{*}{} & & {Sequence length} & 10 \\ 
    \multirow{7}{*}{} & & \cellcolor{gray!15}\makecell[c]{Dimension of Embedding} & \cellcolor{gray!15}1×64 \\  
    \multirow{7}{*}{} & & \makecell[c]{Number of Head } & 4 \\  
    \bottomrule[2pt]
    \end{tabular}
   }
    \caption{Hyperparameters of training.}

\end{table}
\subsubsection{Hyper-parameters Settings} We tune the learning rates of all models in $\{1e^{-3}, 2e^{-3}, 5e^{-3}, 1e^{-2}\}$ and select the best performance. As to model-specific hyper-parameters, we select default parameters from open-source code. Additionally, unless otherwise specified, the length of the user sequence is 10.
All models are trained using a single RTX-4090 GPU. Other hyperparameters of training are shown in table \ref{tab:hyperparameters_training} and the number of head is for transformer-based models.

\subsection{\textbf{Analysis of Resource Consumption(RQ1)}} 
\begin{table}[ht]
\small  
\setlength{\tabcolsep}{8pt}  
\renewcommand{\arraystretch}{1.1}  
  \centering

  \label{tab:param}
  \resizebox{0.45\textwidth}{!}{%
  \begin{tabular}{l|c|c|c}  
    \toprule[2pt]
    \multicolumn{1}{l|}{\textbf{Model}}  
    & \multicolumn{1}{c|}{\textbf{\#Parameter}} 
    & \multicolumn{1}{c|}{\textbf{$\triangle$FLOPs}} 
    & \multicolumn{1}{c}{\textbf{$\triangle$Memory}} \\ 
    \midrule
    \midrule
    DIN & $1.97M$ & $12.6M$  & $2.25M$\\ 
    GRU4Rec & $1.97M$ &  $30.2M$ & $8.78M$\\  
    SASRec & $3.91M$ &  $38.3M$ & $6.01M$\\ 
    BERT4Rec & $2.02M$ &  $101.0M$ & $35.6M$\\ 
    Mamba4Rec & $2.06M$ &  $68.4M$ & $21.6M$\\ 
    \hline 
    \cellcolor{blue!5}CoCorrRec (W/o GCN) & \cellcolor{blue!5}$1.95M$ &  \cellcolor{blue!5}$17.0M$ & \cellcolor{blue!5}$14.2M$\\
    \cellcolor{blue!5}CoCorrRec & \cellcolor{blue!5}$2.08M$ &  \cellcolor{blue!5}$82.7M$ & \cellcolor{blue!5}$23.2M$\\ 
    \bottomrule[2pt]
\end{tabular}}
  \caption{Resource consumption Comparison. $\triangle x$ represents the growth per token of x. }

\end{table}
\begin{figure}
    \centering
    \includegraphics[width=1\linewidth]{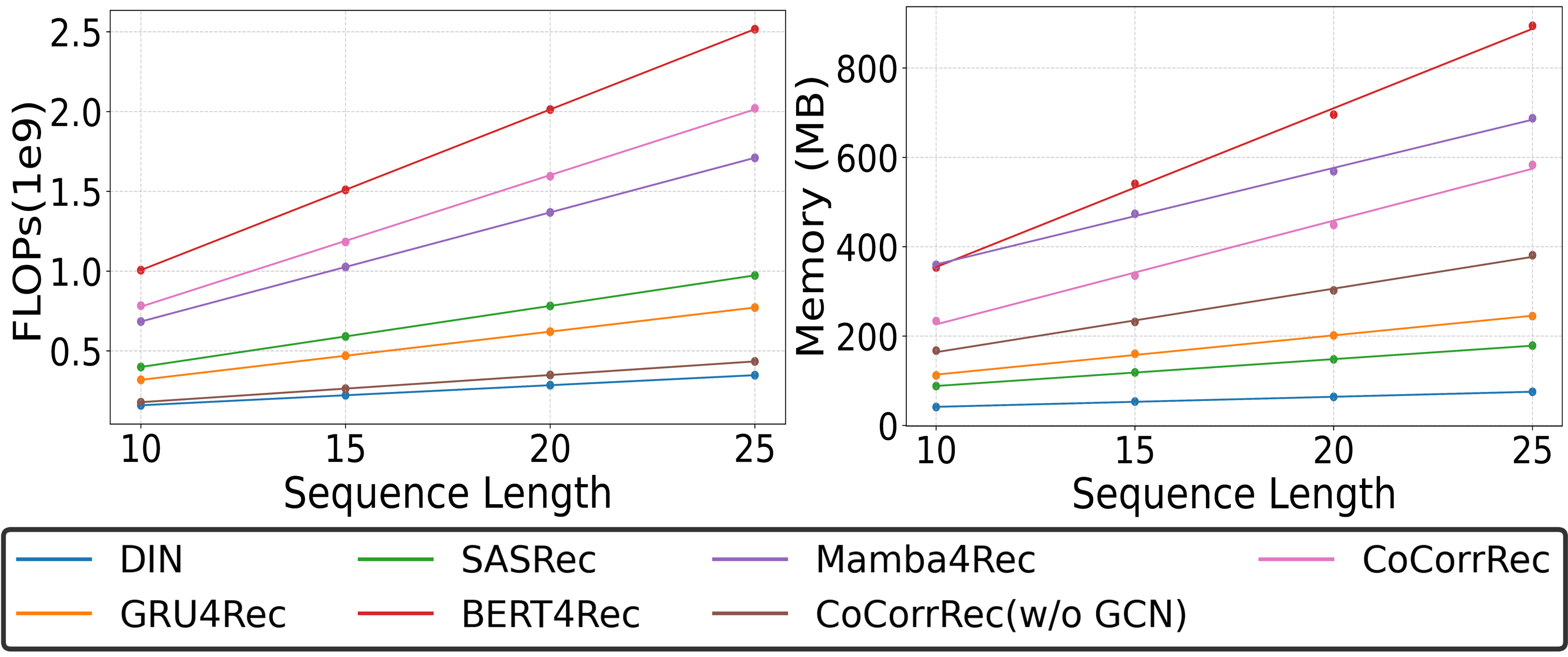}
    \caption{The FLOPs and Max-memory usage of models as the sequence length increasing.}
    \label{fig:flops_memory}
\end{figure}
\begin{table*}[t]
\small

  \label{tab:delivery_delay}
  \centering
  {
    \begin{tabular}{|c|c|c|c|c|c|c|}
      \hline
      \textbf{Datasets}  & \textbf{Size} & \textbf{4G: 5MB/s}  & \textbf{4G: 15MB/s} & \textbf{5G: 50MB/s} & \textbf{5G: 100MB/s} & \textbf{Tolerance} \\
      \hline
      \multirow{3}{*}{\makecell[c]{\texttt{Beauty}}} & \multirow{3}{*}
      {\makecell{$\Uparrow$: 2.56KB \\ $\circ$ : 1.18ms\\ {$\Downarrow$: 4.16KB}}} & 
      \multirow{3}{*}{\makecell{$\Uparrow$: 0.51ms\\ $\circ$ : 1.18ms\\ {$\Downarrow$: 0.83ms}}$\rightarrow$ 2.52ms} & 
      \multirow{3}{*}{\makecell{$\Uparrow$: 0.17ms\\ $\circ$ : 1.18ms\\ {$\Downarrow$: 0.28ms}}$\rightarrow$ 1.63ms} & 
      \multirow{3}{*}{\makecell{$\Uparrow$: 0.051ms\\ $\circ$ : 1.18ms\\ {$\Downarrow$: 0.083ms}}$\rightarrow$ 1.32ms} & 
      \multirow{3}{*}{\makecell{$\Uparrow$: 0.026ms\\ $\circ$ : 1.18ms\\ {$\Downarrow$: 0.042ms}}$\rightarrow$ 1.25ms} & 
      \multirow{3}{*}{\makecell{4.21ms}} \\ 
      & & & & & & \\ 
      & & & & & & \\ 
      \hline
    \end{tabular}
  }
  \caption{Time delay caused by device-cloud communication, $\Uparrow$ and $\Downarrow$ represent upload (device $\rightarrow$ cloud) and download (cloud $\rightarrow$ device), respectively. \textbf{$\circ$} indicates time required to generate correction on cloud.}

\end{table*}
On resource-constrained devices, the increasing FLOPs (Floating Point Operations Per Second) and memory consumption of models as the user interaction sequence grows during the recommendation process can gradually degrade the user experience. Table \ref{tab:param} details the parameters and resource consumption of each model. Notably, most of these models have a comparable number of parameters. The FLOPs metric quantifies the computational cost of a single inference, while memory consumption indicates the peak memory usage during a complete inference epoch on Amazon-Beauty. Figure \ref{fig:flops_memory} illustrates the FLOPs and maximum memory usage of the models when the user sequence lengths are 10, 15, 20, and 25, respectively. 

It is expected that RNN-based models grow linearly with sequence length, but it is surprising that Transformer-based models like SASRec exhibit the same behavior. In fact, for Transformers, the total computation of self-attention (SA) is \(4nh^2d^2 + 2n^2hd\), where \(d\) is the head size and \(h\) is the number of heads, while the Feed-Forward Network (FFN) computation is \(8nh^2d^2\). Therefore, for a Transformer-base (\(d=64\), \(h=12\)), when the sequence length $n \leq 1536$, the linear complexity of the FFN dominates, which is the usual circumstance for recommendation.

It can be observed that our designed CoCorrRec (w/o. GCN) has certain advantages in terms of computational cost and memory usage, with minimal additional overhead as the user sequence grows. Under the same resource constraints, it can handle longer user interaction sequences. Table \ref{tab:param} also shows the additional cost introduced by the GCN. Although the number of additional parameters is small, the FLOPs and memory consumption of the computationally intensive GCN are significant. Therefore, it is more suitable for deployment on cloud with abundant computational resources.
\subsection{Analysis of Time Delay(RQ2)}
\label{sec:time_delay}
\begin{figure}
    \centering
    \includegraphics[width=1\linewidth]{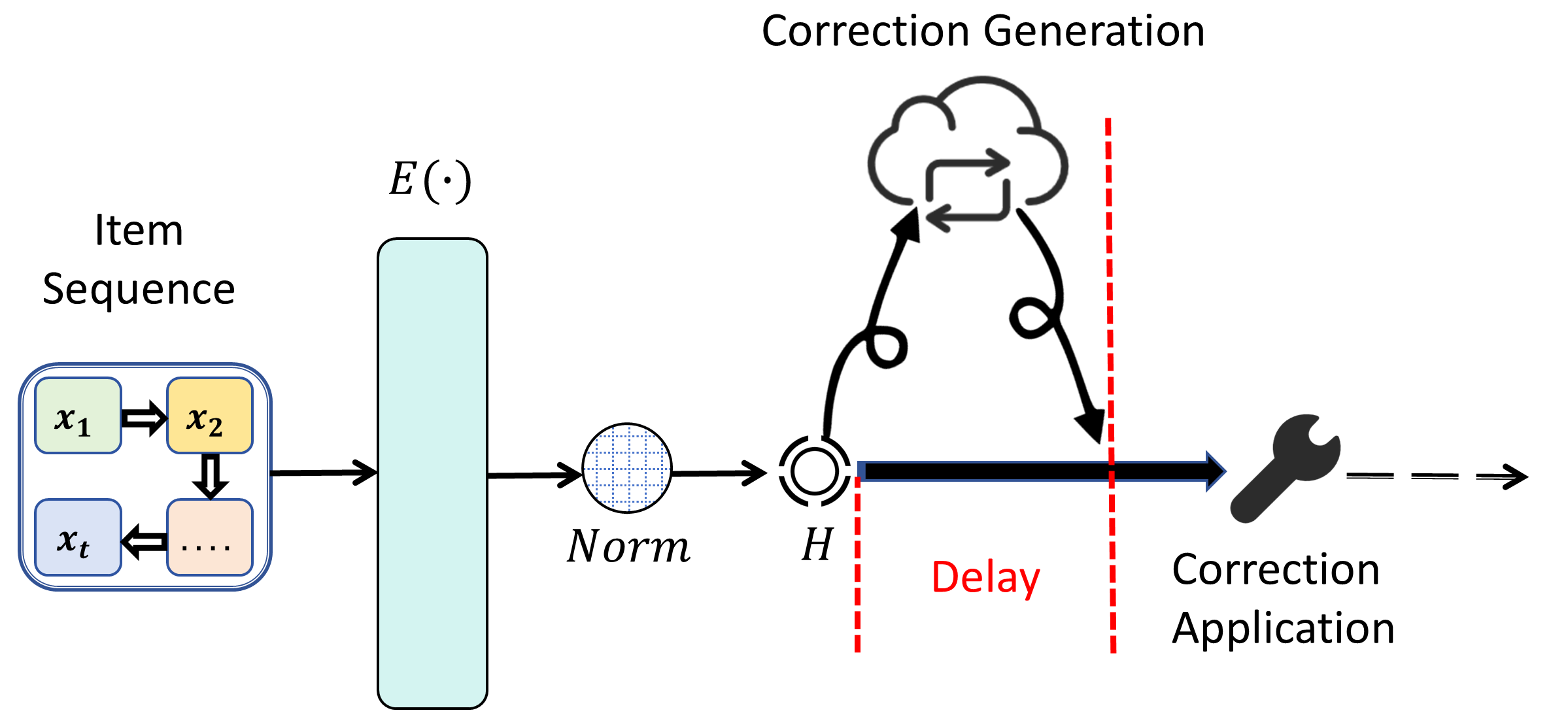}
    \caption{Illustration of the delay caused by cloud-device communication.}
    \label{fig:delay}
\end{figure}
Table \ref{tab:delivery_delay} summarizes the size and time delay involved in uploading hidden state $H$ and downloading correction parameters to the device. It also details the time required for the cloud-based GCN to generate corrections and the device's tolerance time (i.e., the duration from hidden state $H$ generation to the application of corrections on the device) as shown in figure \ref{fig:delay}. The results demonstrate that the time needed for on-device model to receive corrections consistently remains within the tolerance limit, confirming its suitability for real-time applications. Furthermore, uploading embeddings for real-time samples enhances user privacy. These findings validate that the cloud-based GCN does not introduce additional delays to the on-device model, as both processes can run concurrently. This result highlights the effectiveness of CoCorrRec in achieving real-time application requirements.

\subsection{Analysis of CoCorrRec (RQ3)}
\subsubsection{Overall Performance}

Table 1 compares the performance of CoCorrRec with baseline models. CoCorrRec (w/o. GCN) represents the pure on-device model, while CoCorrRec integrates the on-device model with the cloud-based GCN. As this study focuses on sequential recommendation tasks, all models were exclusively trained and tested on sequential recommendation datasets. From Table \ref{tab:main_table}, the following conclusions can be drawn:
\begin{itemize}
\item The SCN demonstrates superior robustness in sequential recommendation tasks. While the performance of various baselines varies across datasets due to their respective strengths, CoCorrRec (w/o. GCN) maintains relatively stable performance across all datasets by updating model parameters during inference.
\item In all cases, our revised CoCorrRec consistently outperforms CoCorrRec (w/o. GCN). This underscores the effectiveness of the cloud-based GCN in enhancing the representation capabilities of the on-device model. Furthermore, the cloud-based GCN improves model performance in parallel without increasing the on-device model’s time overhead.
\end{itemize}
\subsubsection{Effect of Mini-Batch Size}
\begin{figure}
    \centering
    \includegraphics[width=1\linewidth]{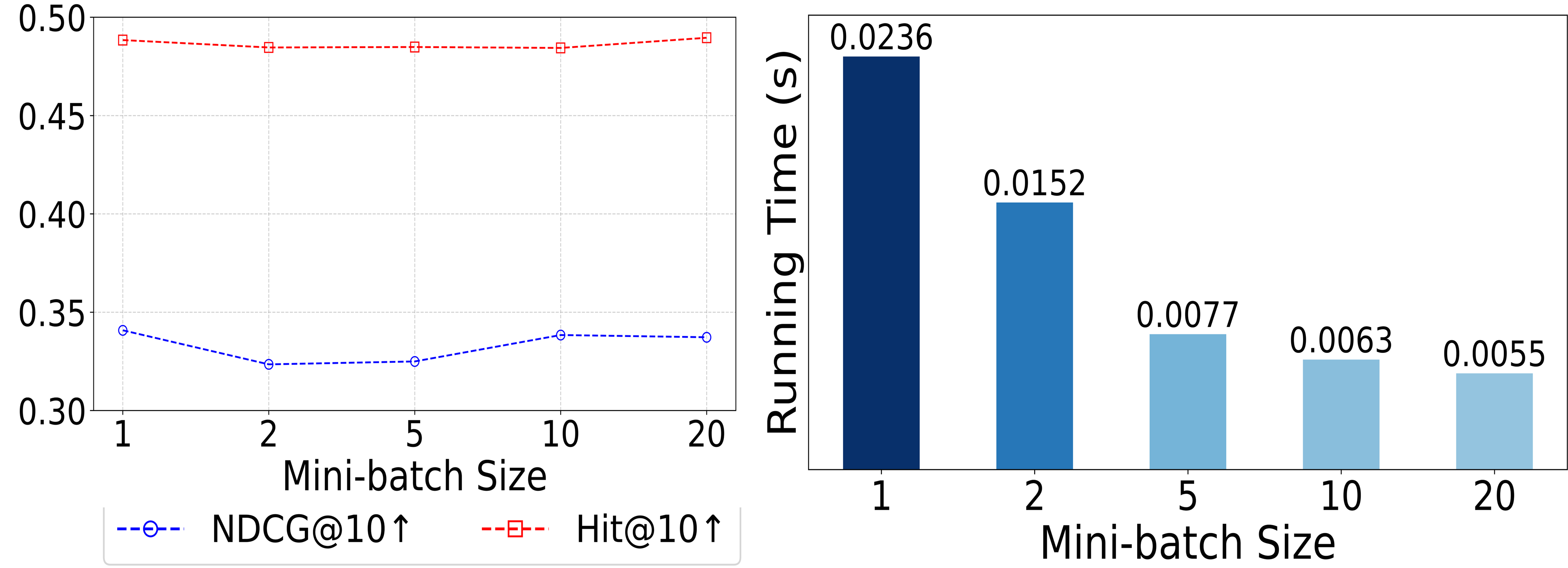}
    \caption{Model performance and inference time with different mini-batch size $b$ on Amazon-Beauty.}
    \label{fig:mini_batch}
\end{figure}
The mini-batch size determines the number of tokens processed simultaneously using the mini-batch parallelization strategy. Although processing all tokens at once minimizes resource consumption, it may not achieve optimal performance. To address this, we assessed the performance of mini-batch sizes of 1, 2, 5, 10, and 20 on the Beauty dataset, with a sequence length of 20. Figure \ref{fig:mini_batch} illustrates the variations in model performance and inference time as the mini-batch size $b$ increases. While performance shows some fluctuation, the results indicate that the performance at $b=20$ is nearly identical to that at $b=1$. However, the inference time at $b=20$ demonstrates a significant improvement. Therefore, parallel computation of all tokens is generally a more efficient choice.
\section{Conclusion}
In this paper, we propose the CoCorrRec, which deploys an RNN-based Self-Correcting Network (SCN) on the device and a Global Correcting Network (GCN) on the cloud. Through the collaboration of both, the model addresses the current limitations of recommendation models on devices in terms of balancing real-time performance and high efficiency. CoCorrRec repositions RNN-based models as highly competitive for sequential recommendation, especially on devices, and demonstrates the potential of device-cloud collaborative parameter correction for device-based recommendation.
\clearpage
\section*{Acknowledgements}
This work was in part supported by 2030 National Science and Technology Major Project (2022ZD0119100), National Natural Science Foundation of China (No. 62402429, U24A20326, 62441605), the Key Research and Development Program of Zhejiang Province (No. 2024C03270), ZJU Kunpeng\&Ascend Center of Excellence, Ningbo Yongjiang Talent Introduction Programme (2023A-397-G), Young Elite Scientists Sponsorship Program by CAST (2024QNRC001).
\bibliographystyle{named}
\bibliography{main}

\begin{thebibliography}{}

\bibitem[\protect\citeauthoryear{Fu \bgroup \em et al.\egroup }{2024}]{fu2024diet}
Kairui Fu, Shengyu Zhang, Zheqi Lv, Jingyuan Chen, and Jiwei Li.
\newblock Diet: Customized slimming for incompatible networks in sequential recommendation.
\newblock In {\em Proceedings of the 30th ACM SIGKDD Conference on Knowledge Discovery and Data Mining}, pages 816--826, 2024.

\bibitem[\protect\citeauthoryear{Fu \bgroup \em et al.\egroup }{2025}]{fu2025forward}
Kairui Fu, Zheqi Lv, Shengyu Zhang, Fan Wu, and Kun Kuang.
\newblock Forward once for all: Structural parameterized adaptation for efficient cloud-coordinated on-device recommendation.
\newblock In {\em Proceedings of the 31st ACM SIGKDD Conference on Knowledge Discovery and Data Mining V.1}, KDD '25, page 318–329, New York, NY, USA, 2025. Association for Computing Machinery.

\bibitem[\protect\citeauthoryear{Gandelsman \bgroup \em et al.\egroup }{2022}]{ref:ttt-mae}
Yossi Gandelsman, Yu~Sun, Xinlei Chen, and Alexei~A. Efros.
\newblock Test-time training with masked autoencoders.
\newblock {\em Advances in Neural Information Processing Systems}, 2022.

\bibitem[\protect\citeauthoryear{Gong \bgroup \em et al.\egroup }{2020}]{gong2020edgerec}
Yu~Gong, Ziwen Jiang, Yufei Feng, Binbin Hu, Kaiqi Zhao, Qingwen Liu, and Wenwu Ou.
\newblock Edgerec: recommender system on edge in mobile taobao.
\newblock In {\em Proceedings of the 29th ACM International Conference on Information \& Knowledge Management}, pages 2477--2484, 2020.

\bibitem[\protect\citeauthoryear{Hardt and Sun}{2023}]{ref:test_LLM}
Moritz Hardt and Yu~Sun.
\newblock Test-time training on nearest neighbors for large language models.
\newblock {\em arXiv preprint arXiv:2305.18466}, 2023.

\bibitem[\protect\citeauthoryear{Hidasi \bgroup \em et al.\egroup }{2016}]{ref:gru4rec}
Bal{\'a}zs Hidasi, Alexandros Karatzoglou, Linas Baltrunas, and Domonkos Tikk.
\newblock Session-based recommendations with recurrent neural networks.
\newblock {\em International Conference on Learning Representations 2016}, 2016.

\bibitem[\protect\citeauthoryear{Kang and McAuley}{2018}]{ref:sasrec}
Wang-Cheng Kang and Julian McAuley.
\newblock Self-attentive sequential recommendation.
\newblock In {\em 2018 IEEE International Conference on Data Mining (ICDM)}, pages 197--206. IEEE, 2018.

\bibitem[\protect\citeauthoryear{Lai \bgroup \em et al.\egroup }{2023}]{ref:SAMRec}
Vivian Lai, Huiyuan Chen, Chin-Chia~Michael Yeh, Minghua Xu, Yiwei Cai, and Hao Yang.
\newblock Enhancing transformers without self-supervised learning: A loss landscape perspective in sequential recommendation.
\newblock In {\em Proceedings of the 17th ACM Conference on Recommender Systems}, RecSys '23, pages 791--797, 2023.

\bibitem[\protect\citeauthoryear{Li \bgroup \em et al.\egroup }{2021}]{ref:Fairness}
Yunqi Li, Hanxiong Chen, Zuohui Fu, Yingqiang Ge, and Yongfeng Zhang.
\newblock User-oriented fairness in recommendation.
\newblock In {\em WWW}, 2021.

\bibitem[\protect\citeauthoryear{Li \bgroup \em et al.\egroup }{2023}]{ref:STRec}
Chengxi Li, Yejing Wang, Qidong Liu, Xiangyu Zhao, Wanyu Wang, Yiqi Wang, Lixin Zou, Wenqi Fan, and Qing Li.
\newblock Strec: Sparse transformer for sequential recommendations.
\newblock In {\em Proceedings of the 17th ACM Conference on Recommender Systems}, RecSys '23, pages 101--111, 2023.

\bibitem[\protect\citeauthoryear{Li \bgroup \em et al.\egroup }{2024}]{Li2024MergeNetKM}
Kunxi Li, Tianyu Zhan, Shengyu Zhang, Kun Kuang, Jiwei Li, Zhou Zhao, and Fei Wu.
\newblock Mergenet: Knowledge migration across heterogeneous models, tasks, and modalities.
\newblock In {\em AAAI Conference on Artificial Intelligence}, 2024.

\bibitem[\protect\citeauthoryear{Lin \bgroup \em et al.\egroup }{2018}]{ref:KrNN-P}
Xiang Lin, Shuzi Niu, Yiqiao Wang, and Yucheng Li.
\newblock K-plet recurrent neural networks for sequential recommendation.
\newblock In {\em The 41st International ACM SIGIR Conference on Research and Development in Information Retrieval}, SIGIR '18, pages 1057--1060, 2018.

\bibitem[\protect\citeauthoryear{Liu \bgroup \em et al.\egroup }{2023}]{ref:AutoSeqRec}
Sijia Liu, Jiahao Liu, Hansu Gu, Dongsheng Li, Tun Lu, Peng Zhang, and Ning Gu.
\newblock Autoseqrec: Autoencoder for efficient sequential recommendation.
\newblock In {\em Proceedings of the 32nd ACM International Conference on Information and Knowledge Management}, CIKM '23, pages 1493--1502, 2023.

\bibitem[\protect\citeauthoryear{Liu \bgroup \em et al.\egroup }{2024}]{ref:mamba4rec}
Chengkai Liu, Jianghao Lin, Jianling Wang, Hanzhou Liu, and James Caverlee.
\newblock Mamba4rec: Towards efficient sequential recommendation with selective state space models, 2024.

\bibitem[\protect\citeauthoryear{Long \bgroup \em et al.\egroup }{2024a}]{ref:MLM4Rec}
Chao Long, Huanhuan Yuan, Junhua Fang, Xuefeng Xian, Guanfeng Liu, Victor~S Sheng, and Pengpeng Zhao.
\newblock Learning global and multi-granularity local representation with mlp for sequential recommendation.
\newblock {\em ACM Transactions on Knowledge Discovery from Data}, 18(4):1--15, 2024.

\bibitem[\protect\citeauthoryear{Long \bgroup \em et al.\egroup }{2024b}]{ref:device_cloud_rec}
Jing Long, Guanhua Ye, Tong Chen, Yang Wang, Meng Wang, and Hongzhi Yin.
\newblock Diffusion-based cloud-edge-device collaborative learning for next {POI} recommendations.
\newblock In {\em {KDD}}, pages 2026--2036. {ACM}, 2024.

\bibitem[\protect\citeauthoryear{Lv \bgroup \em et al.\egroup }{2023}]{ref:duet}
Zheqi Lv, Wenqiao Zhang, Shengyu Zhang, Kun Kuang, Feng Wang, Yongwei Wang, Zhengyu Chen, Tao Shen, Hongxia Yang, Beng~Chin Ooi, and Fei Wu.
\newblock {DUET:} {A} tuning-free device-cloud collaborative parameters generation framework for efficient device model generalization.
\newblock In {\em {WWW}}, pages 3077--3085. {ACM}, 2023.

\bibitem[\protect\citeauthoryear{Lv \bgroup \em et al.\egroup }{2024}]{ref:intelligent}
Zheqi Lv, Wenqiao Zhang, Zhengyu Chen, Shengyu Zhang, and Kun Kuang.
\newblock Intelligent model update strategy for sequential recommendation.
\newblock In {\em {WWW}}, pages 3117--3128. {ACM}, 2024.

\bibitem[\protect\citeauthoryear{Lv \bgroup \em et al.\egroup }{2025}]{lv2025collaboration}
Zheqi Lv, Tianyu Zhan, Wenjie Wang, Xinyu Lin, Shengyu Zhang, Wenqiao Zhang, Jiwei Li, Kun Kuang, and Fei Wu.
\newblock Collaboration of large language models and small recommendation models for device-cloud recommendation.
\newblock In {\em {KDD} {(1)}}, pages 962--973. {ACM}, 2025.

\bibitem[\protect\citeauthoryear{Muhammad \bgroup \em et al.\egroup }{2020}]{ref:fedfast}
Khalil Muhammad, Qinqin Wang, Diarmuid O'Reilly-Morgan, Elias Tragos, Barry Smyth, Neil Hurley, James Geraci, and Aonghus Lawlor.
\newblock Fedfast: Going beyond average for faster training of federated recommender systems.
\newblock In {\em KDD}, 2020.

\bibitem[\protect\citeauthoryear{Niu and Zhang}{2017}]{ref:MrRNN}
Shuzi Niu and Rongzhi Zhang.
\newblock Collaborative sequence prediction for sequential recommender.
\newblock In {\em Proceedings of the 2017 ACM on Conference on Information and Knowledge Management}, pages 2239--2242, 2017.

\bibitem[\protect\citeauthoryear{Peng \bgroup \em et al.\egroup }{2023}]{ref:MSSG}
Bo~Peng, Ziqi Chen, Srinivasan Parthasarathy, and Xia Ning.
\newblock Modeling sequences as star graphs to address over-smoothing in self-attentive sequential recommendation.
\newblock {\em ACM Transactions on Knowledge Discovery from Data}, 2023.

\bibitem[\protect\citeauthoryear{Petrov and Macdonald}{2022}]{ref:ALBERT4Rec_DeBERTa4Rec}
Aleksandr Petrov and Craig Macdonald.
\newblock A systematic review and replicability study of bert4rec for sequential recommendation.
\newblock In {\em Proceedings of the 16th ACM Conference on Recommender Systems}, RecSys '22, pages 436--447, 2022.

\bibitem[\protect\citeauthoryear{Qian \bgroup \em et al.\egroup }{2022}]{ref:device_cloud_adarequest}
Xufeng Qian, Yue Xu, Fuyu Lv, Shengyu Zhang, Ziwen Jiang, Qingwen Liu, Xiaoyi Zeng, Tat-Seng Chua, and Fei Wu.
\newblock Intelligent request strategy design in recommender system.
\newblock In {\em Proceedings of the 28th ACM SIGKDD Conference on Knowledge Discovery and Data Mining}, pages 3772--3782, 2022.

\bibitem[\protect\citeauthoryear{Rendle \bgroup \em et al.\egroup }{2010}]{ref:FPMC}
Steffen Rendle, Christoph Freudenthaler, and Lars Schmidt-Thieme.
\newblock Factorizing personalized markov chains for next-basket recommendation.
\newblock In {\em Proceedings of the 19th International Conference on World Wide Web}, TheWebConf '10, pages 811--820, 2010.

\bibitem[\protect\citeauthoryear{Su \bgroup \em et al.\egroup }{2023a}]{su2023personalized}
Jiajie Su, Chaochao Chen, Zibin Lin, Xi~Li, Weiming Liu, and Xiaolin Zheng.
\newblock Personalized behavior-aware transformer for multi-behavior sequential recommendation.
\newblock In {\em Proceedings of the 31st ACM international conference on multimedia}, pages 6321--6331, 2023.

\bibitem[\protect\citeauthoryear{Su \bgroup \em et al.\egroup }{2023b}]{su2023enhancing}
Jiajie Su, Chaochao Chen, Weiming Liu, Fei Wu, Xiaolin Zheng, and Haoming Lyu.
\newblock Enhancing hierarchy-aware graph networks with deep dual clustering for session-based recommendation.
\newblock In {\em Proceedings of the ACM web conference 2023}, pages 165--176, 2023.

\bibitem[\protect\citeauthoryear{Su \bgroup \em et al.\egroup }{2024}]{su2024roformer}
Jianlin Su, Murtadha Ahmed, Yu~Lu, Shengfeng Pan, Wen Bo, and Yunfeng Liu.
\newblock Roformer: Enhanced transformer with rotary position embedding.
\newblock {\em Neurocomputing}, 568:127063, 2024.

\bibitem[\protect\citeauthoryear{Sun \bgroup \em et al.\egroup }{2019}]{ref:bert4rec}
Fei Sun, Jun Liu, Jian Wu, Changhua Pei, Xiao Lin, Wenwu Ou, and Peng Jiang.
\newblock Bert4rec: Sequential recommendation with bidirectional encoder representations from transformer.
\newblock In {\em Proceedings of the 28th ACM international conference on information and knowledge management}, pages 1441--1450, 2019.

\bibitem[\protect\citeauthoryear{Sun \bgroup \em et al.\egroup }{2020}]{ref:test}
Yu~Sun, Xiaolong Wang, Zhuang Liu, John Miller, Alexei Efros, and Moritz Hardt.
\newblock Test-time training with self-supervision for generalization under distribution shifts.
\newblock In {\em International Conference on Machine Learning}, pages 9229--9248. PMLR, 2020.

\bibitem[\protect\citeauthoryear{Sun \bgroup \em et al.\egroup }{2024}]{ref:TTT}
Yu~Sun, Xinhao Li, Karan Dalal, Jiarui Xu, Arjun Vikram, Genghan Zhang, Yann Dubois, Xinlei Chen, Xiaolong Wang, Sanmi Koyejo, Tatsunori Hashimoto, and Carlos Guestrin.
\newblock Learning to (learn at test time): Rnns with expressive hidden states, 2024.

\bibitem[\protect\citeauthoryear{Tang and Wang}{2018}]{ref:Caser}
Jiaxi Tang and Ke~Wang.
\newblock Personalized top-n sequential recommendation via convolutional sequence embedding.
\newblock In {\em Proceedings of the 11th ACM International Conference on Web Search and Data Mining}, WSDM '18, pages 565--573, 2018.

\bibitem[\protect\citeauthoryear{Wang \bgroup \em et al.\egroup }{2023a}]{ref:test_video}
Renhao Wang, Yu~Sun, Yossi Gandelsman, Xinlei Chen, Alexei~A Efros, and Xiaolong Wang.
\newblock Test-time training on video streams.
\newblock {\em arXiv preprint arXiv:2307.05014}, 2023.

\bibitem[\protect\citeauthoryear{Wang \bgroup \em et al.\egroup }{2023b}]{ref:diffusion4rec}
Wenjie Wang, Yiyan Xu, Fuli Feng, Xinyu Lin, Xiangnan He, and Tat-Seng Chua.
\newblock Diffusion recommender model, 2023.

\bibitem[\protect\citeauthoryear{Yan \bgroup \em et al.\egroup }{2022}]{ref:device_cloud}
Yikai Yan, Chaoyue Niu, Renjie Gu, Fan Wu, Shaojie Tang, Lifeng Hua, Chengfei Lyu, and Guihai Chen.
\newblock On-device learning for model personalization with large-scale cloud-coordinated domain adaption.
\newblock In {\em {KDD} '22: The 28th {ACM} {SIGKDD} Conference on Knowledge Discovery and Data Mining, Washington, DC, USA, August 14 - 18, 2022}, pages 2180--2190, 2022.

\bibitem[\protect\citeauthoryear{Yao \bgroup \em et al.\egroup }{2021}]{ref:dccl}
Jiangchao Yao, Feng Wang, Kunyang Jia, Bo~Han, Jingren Zhou, and Hongxia Yang.
\newblock Device-cloud collaborative learning for recommendation.
\newblock In {\em {KDD}}, pages 3865--3874. {ACM}, 2021.

\bibitem[\protect\citeauthoryear{Yao \bgroup \em et al.\egroup }{2022}]{ref:MetaC}
Jiangchao Yao, Feng Wang, Xichen Ding, Shaohu Chen, Bo~Han, Jingren Zhou, and Hongxia Yang.
\newblock Device-cloud collaborative recommendation via meta controller.
\newblock In {\em {KDD}}, pages 4353--4362. {ACM}, 2022.

\bibitem[\protect\citeauthoryear{Zhan \bgroup \em et al.\egroup }{2025}]{zhan2025preliminary}
Tianyu Zhan, Zheqi Lv, Shengyu Zhang, and Jiwei Li.
\newblock Preliminary evaluation of the test-time training layers in recommendation system (student abstract).
\newblock In {\em Proceedings of the AAAI Conference on Artificial Intelligence}, volume~39, pages 29554--29557, 2025.

\bibitem[\protect\citeauthoryear{Zhang \bgroup \em et al.\egroup }{2020}]{zhang2020devlbert}
Shengyu Zhang, Tan Jiang, Tan Wang, Kun Kuang, Zhou Zhao, Jianke Zhu, Jin Yu, Hongxia Yang, and Fei Wu.
\newblock Devlbert: Learning deconfounded visio-linguistic representations.
\newblock In {\em Proceedings of the 28th ACM International Conference on Multimedia}, pages 4373--4382, 2020.

\bibitem[\protect\citeauthoryear{Zhou \bgroup \em et al.\egroup }{2018}]{ref:din}
Guorui Zhou, Xiaoqiang Zhu, Chenru Song, Ying Fan, Han Zhu, Xiao Ma, Yanghui Yan, Junqi Jin, Han Li, and Kun Gai.
\newblock Deep interest network for click-through rate prediction.
\newblock In {\em Proceedings of the 24th ACM SIGKDD International Conference on Knowledge Discovery \& Data Mining}, pages 1059--1068, 2018.

\bibitem[\protect\citeauthoryear{Zhou \bgroup \em et al.\egroup }{2022}]{ref:FMLP-Rec}
Kun Zhou, Hui Yu, Wayne~Xin Zhao, and Ji-Rong Wen.
\newblock Filter-enhanced mlp is all you need for sequential recommendation.
\newblock In {\em Proceedings of the ACM Web Conference 2022}, TheWebConf '22, pages 2388--2399, 2022.

\end{thebibliography}
\end{document}